# Robust intrinsic electronic superconducting phases in underdoped $La_{2-x}Sr_xCuO_4$ single crystals

X.L. Dong1, 2, P.H. Hor1∗, F. Zhou2 and Z.-X. Zhao2

*1 Texas Center for Superconductivity and Department of Physics, University of Houston, Houston, TX 77204-5002, USA*

*2 Beijing National Laboratory for Condensed Matter Physics, Institute of Physics, Chinese Academy of Sciences, Beijing 100080, China*

We have measured the superconducting critical temperature ($T_C$) and the diamagnetic susceptibility of $La_{2-x}Sr_xCuO_4$ single crystals in various magnetic fields. We observed a field-induced evolution from an apparent $T_C$ phase to an intrinsic $T_{C1}$ = 15 K or $T_{C2}$ = 30 K phase characterized by "magic" hole concentration which is commensurate with that of a two dimensional electronic lattice. The onset $T_C$ of the intrinsic superconducting phases remains robust up to H = 5 Tesla. We suggest that the intrinsic superconducting phases at "magic" doping concentrations are the pristine electronic phases of high temperature superconductivity.

---

∗ Corresponding author, email: phor@uh.edu, fax: 1-713-743-8301

It has become clear in recent years that the inhomogeneity is one of the salient features in high temperature superconductors (HTS). To pin down the driving forces of the underlying inhomogeneities is critical to the ultimate construction of the microscopic theory of HTS. Since HTS are derived from chemically doping holes into the parent Mott insulating cuprates there are two types of intrinsic electronic inhomogeneities: (1) dopant-induced electronic random disorder [1] and (2) the development of novel electronic phases such as one-dimensional stripes [2] or two-dimensional (2D) square-lattice charge ordering [3,4]. The existence of electronic phases naturally involves electronic phase separations. Indeed, the electronic phase separation (EPS) into an antiferromagnetic phase and a spin-glass phase was observed in Sr-doped $La_{2-x}Sr_xCuO_4$ (SD-La214) for $x < 0.02$ [5]. In the underdoped regime, there is an EPS into two intrinsic $T_{C1} = 15$ K and $T_{C2} = 30$ K superconducting phases in the Sr/O co-doped $La_{2-x}Sr_xCuO_{4+\delta}$ (CD-La214) [6]. Far infrared (FIR) charge dynamics studies suggested that, independent of the nature of dopants, $T_{C1}$ and $T_{C2}$ are very peculiar electronic phases where only very small amount (< 1% of the total doped holes) of free carriers are moving in the otherwise pinned 2D electronic lattices (EL) with p(4 x 4) and c(2 x 2) symmetries at "magic" planar hole concentrations of $x = 1/16$ and $2/16$, respectively [3]. Note that these 2D EL's exist even at room temperature and continuously develop as temperature decreases [3, 4]. Most recently real space observation of 4a x 4a 2D charge ordering was reported in Bi2212 crystal [7] and rational doping fractions consistent with checkerboard-type ordering of Cooper pairs were reported in SD-La214 crystals [8].

The intrinsic $T_C$ phases are illusive: they are energetically close and greatly affected by random disorder induced by dopants such that nanoscale EPS always results in Sr-doped samples, even in the high quality single crystals purposely grown at magic doping concentrations [4]. Here we use magnetic field to sort out and study the intrinsic $T_C$ phases in SD-La214 single crystals. The magnetic field probe is gentle and effective. It can "probe" the superconducting state without introducing extra disordering effect due to chemical dopants [9]. Since $T_{C1}$ and $T_{C2}$ are energetically favored electronic phases it is expected that, under external perturbations such as applying pressure or magnetic field, the superconductor will settle into a globally optimized state. This optimal state, characterized by the underlying electronic texture, can vary from microscopic to macroscopic mixtures of the intrinsic electronic phases depending on whether it is hard-doping or soft-doping, respectively [6]. In this Letter, we report a surprising result that the onset temperatures of $T_{C1}$ and $T_{C2}$ are 'robust' under externally applied magnetic field up to 5 Tesla. Therefore, upon applying magnetic fields, all non-intrinsic $T_C$'s have eventually transformed and settled into a single $T_C$ that equals to either $T_{C1}$ or $T_{C2}$ for H > 1 Tesla. Furthermore, the reorganizing behavior is also observed in the plot of the signal size of FC (field-cooled) diamagnetic susceptibility at 5K ($^{5K}\chi_{FC} = {}^{5K}M_{FC}/H$) as a function of applied magnetic field.

Five crystals with hole-doping levels x = 0.063 (~1/16), 0.07, 0.09, 0.11 (~1/9) and 0.125 (=2/16) and $T_C$ ranging from 15 K to 37 K were processed into approximately 1 x 2 x 0.3 mm$^3$ rectangular shape with the c-axis perpendicular to the largest surface. Details of the crystal growth and characterization were reported previously [10]. We measured the magnetic susceptibilities with applied field (H) parallel to the c-axis using Quantum Design SQUID magnetometer MPMS-XL (for H < 1 Tesla) and MPMS-5S (for H > 1 Tesla). The MPMS-XL has a monotonic field profile with high overall uniformity (< 1 mOe for H < 10 Oe and ~ 0.25% for H up to 1 Tesla) and small residual H (< 0.5 Oe) when using 4 cm scan length. To minimize the flux trapping caused by H-inhomogeneity or temperature (T) change, the magnetization was always measured during warming cycle under a fixed H and the sample was always positioned at the center of the detecting coil during cooling or adjusting H. All the data reported have been corrected for the demagnetization factor [11].

In Fig. 1, we show differential FC diamagnetic susceptibility ($\chi$) data measured at 5 Oe. There are extremely sharp single transitions of $T_{C1}$ = 15 K and $T_{C2}$ = 30 K for x ~ 0.063 and ~ 0.11 [12] samples, respectively. However, in the same series of high quality crystals, whenever the $T_C$ is away from $T_{C1}$ or $T_{C2}$ the superconducting transition becomes much broader, manifested as broad $d\chi/dT$ curves with multiple-peaks structure. These behaviors are consistent with and can be understood in terms of the existence of the electronic inhomogeneity due to nano-scale mixing of the 2D EL's [3, 4, 6, 10].

In Fig. 2 we show that the intrinsic $T_C$'s are "robust" under H and behave characteristically different from that of the non-intrinsic ones. The T-dependence of $d\chi_{FC}/dT$ of the x = 0.11 crystal as a function of H clearly shows that while the transition width ($\Delta T_C$) of the intrinsic $T_{C2}$ of x = 0.11 crystal increases with increasing H, the onset [13] of the superconducting transition at 30K is surprisingly 'robust' with little field dependence (Fig.2a). In contrast, as seen in Fig. 2b, the width of broad $d\chi_{FC}/dT$ curve of x = 0.125 sample does not broaden with increasing H, in stead, the initial apparent $T_C$ = 37 K is continuously suppressed until it reaches $T_{C2}$ at H ~ 1 Tesla. Again, $\Delta T_C$ then starts to increase rapidly while $T_{C2}$ exhibits little field dependence for 1 Tesla < H < 5 Tesla. The extra peak-like features that appear in both crystals for H > 0.1 Tesla originate from the hump of the magnetization due to the well-known vortex phase transition in the vicinity of the irreversibility temperature [14, 15].

Summarized by the color map in Fig. 3: we observe the same robustness of $T_{C1}$ (=15 K) in x = 0.063 crystal (Fig. 3a). However whenever $T_C$ is away from 15K, such as that of x = 0.07 crystal ($T_C$ ~ 22K at H = 1Oe), the broad $\Delta T_C$ becomes narrower as $T_C$ decreases with increasing H and then starts to increase at H = 0.5 Tesla as $T_C$ reaches 15K from above (see inset of Fig. 3b). For H ≥ 1 Tesla, the $T_C$ remains at 15K and $\Delta T_C$ increases sharply. Similar behavior is also observed in x = 0.09 crystal (Fig. 3c); the transition exhibits a two-peak structure, one at $T_{C2}$ and the other at 23K. The $\Delta T_C$ of the second transition shrinks and disappears for H ≥ 0.1 Tesla. Concomitantly, the $\Delta T_C$ of $T_{C2}$ broadens upon increasing H with $T_{C2}$ exhibiting little field dependence for 0.1 Tesla < H < 5 Tesla. Therefore, when tuned by magnetic field, the superconducting transitions of x = 0.063 and 0.07 crystals converge to $T_{C1}$ = 15K and that of x = 0.09, 0.11 and 0.125 crystals converge to $T_{C2}$ = 30K. Furthermore the anomalous broadening and narrowing of the $\Delta T_C$ under field are closely tied to the approaching, from above, to the two robust superconducting states through H-tuning. This is characteristically different from that of conventional type II superconductors or granular superconductors where there is only one non-robust superconducting state with monotonic decreasing of $T_C$ under field.

Another surprising observation is that $^{5K}\chi_{FC}$(H)'s also group into two distinct branches corresponding to that of $T_{C1}$ and $T_{C2}$ phases for H > 1 Tesla (see Fig. 4). $^{5K}\chi_{FC}$(H) is a measure of the flux exclusion from a bulk superconductor when cooling in a field H from above $T_C$ to 5K. It is greatly affected by flux pinning [16]. To check this point we measured four additional SD-La214 crystals of various sizes and shapes such that the mass changes by two orders of magnitude ranging from 2.28 mg to 177.2 mg. Indeed, there is no systematic $^{5K}\chi_{FC}$ dependence on the sample size and the doping level for H < 0.1 Tesla, a clear indication of the presence of non-reproducible extrinsic pinning effects even in our high quality single crystals. However, as seen in Fig. 4, $^{5K}\chi_{FC}(H)$'s start to group into two distinct curves for H > 0.1 Tesla. For H > 1 Tesla, $^{5K}\chi_{FC}(H)$'s of x = 0.063 and 0.07 samples have merged into one universal curve while those of x = 0.09, 0.11 and 0.125 samples fall on another universal curve. In fact all the available FC data found in the literatures [14, 15, 17] fall on the universal $^{5K}\chi_{FC}$-H curve of $T_{C2}$ for H > 1 Tesla.

Noted that the "grouping" of $^{5K}\chi_{FC}(H)$'s does not depend on the size, the shape or even the hole concentration of the crystals; it depends only on the bulk $T_{C1}$ or $T_{C2}$. Therefore each universal $^{5K}\chi_{FC}$-H curve is an "intrinsic" property of $T_{C1}$ phase or $T_{C2}$ phase. This indicates that $T_{C1}$ and $T_{C2}$ are indeed genuine thermodynamic phases with its own distinct condensate state that eventually overcomes the extrinsic pinning effects and become dominate for H > 1 Tesla. This is very unusual for a superconductor since increasing magnetic field is effectively increasing disordering effects and in general the diamagnetic susceptibility depends, even without pinning, on the shape and the size of the sample. Therefore we can view the intrinsic $T_{C1}$ and $T_{C2}$ phases as a phase with an effective diamagnetic moment ~ 1.7 x$10^{-4}$ $\mu_B$ per-copper-ion and ~ 5 x$10^{-3}$ $\mu_B$ per-copper-ion at 5K in a 5 Tesla field, respectively.

There are, as seen in Fig. 4, three characteristic field regimes in $^{5K}\chi_{FC}$-H curve for all SD-La 214 crystals studied: (1) the extrinsic pinning dominated regime where there is no systematic change of $^{5K}\chi_{FC}$ with the doping level and sample size for H < 0.1 Tesla; (2) the crossover regime where the condensate state of either $T_{C1}$ or $T_{C2}$ starts to win over pinning for 0.1 Tesla < H < 1 Tesla; and (3) the intrinsic condensate regime where $^{5K}\chi_{FC}$ is dominated by either $T_{C1}$ phase or $T_{C2}$ phase for 1 Tesla < H < 5 Tesla. In contrast to the robust $T_{C1}$ and $T_{C2}$ the apparent non-intrinsic $T_C$ in regime 1 and 2 changes with applied magnetic field.

The static EPS picture proposed by the FIR charge dynamics studies has a spatial resolution which is limited by the wavelength of the FIR light, typically, from 25 μm to 1000 μm. Therefore FIR probe can not detect many reported nanoscale dynamical inhomogeneities in the charge and lattice degrees of freedom [18]. In particular, time-resolved optical experiments revealed a temperature-dependent *universal* length scale for dynamic charge inhomogeneity ($l_e$) in SD-La214 system with $l_e$ ( T = 25K ) ~ 300 Å that continuously increases to $l_e$ ~ 1100 Å when T = 10 K [18]. Since the inter-vortex spacing $l_v = (\Phi_0/B)^{0.5}$, we find that regimes-1, -2, and -3 are actually corresponding to $l_e$ (T = 10K) << $l_v$, $l_e$( T = 25K ) < $l_v$ < $l_e$ ( T = 10K ) and $l_v$ << $l_e$ ( T = 10K ), respectively. Therefore, when cooled from normal state to 5K in a magnetic field, the distinct characteristic regimes sorted out by H were actually dictated by the universal dynamic landscape of the underlying electronic texture. Therefore the two distinct $^{5K}\chi_{FC}$-H curves in Fig. 4 led us to conclude that $l_e$ is due to the thermodynamic fluctuations of the intrinsic $T_{C1}$ and $T_{C2}$ phases. Upon entering the superconducting state the applied magnetic field will "see" the dynamic nanoscale domains of intrinsic phase only when the inter-vortex spacing becomes comparable to that of universal length scale. Consequently, in regime-1 the vortex lattice is very dilute and the bulk superconductivity is realized through, similar to a granular superconductor, globally coupling nanoscale domains of various intrinsic electronic phases with a field-dependent apparent $T_C$; in regime-2 the influences of the intrinsic $T_C$ phase kick in and becomes increasingly dominate with increasing field, hence, the apparent $T_C$'s gradually merge into either $T_{C1}$ or $T_{C2}$ and $^{5K}\chi_{FC}$ start evolving into two groups; in regime-3 the condensate state is dominated by either intrinsic $T_{C1}$ or $T_{C2}$ phase.

Finally it is interesting to point out that a recent neutron scattering experiment has provided a rare glimpse into the $T_{C2}$ phase through the observation of a field-induced three-dimensional (3D) antiferromagnetism in x = 0.10 CD-La214 crystals [19]. It is shown that antiferromagnetism is nucleated by vortices and spatially coexisted with superconductivity. Note that EL is also 3D with finite inter-planar correlations as seen by FIR probe [20]. It indicates that, in regime-3, $T_{C2}$ phase is a very peculiar state where 3D charge ordering at magic doping concentration, 3D antiferromagnetism and superconductivity spatially coexist at different length scales.

In summary, we have identified two, $T_{C1}$ = 15K and $T_{C2}$ = 30K, intrinsic electronic superconducting phases at magic doping concentrations commensurate with that of 2D square lattices in underdoped lanthanum cuprates. They have been observed in samples with vastly different types of dopants (Sr/O co-doping vs. pure Sr-doping), sample conditions (poly-crystal vs. single crystal) and tuning parameters (pressure vs. magnetic field). $T_{C1}$ and $T_{C2}$ are robust under field up to 5 Tesla. There is a distinct superconducting condensate state for each intrinsic superconducting phase. It is noted that field-induced distinct condensate state has also been suggested in the double layer Bi2212 [21] and YBCO [22] systems. We suggest that the intrinsic superconducting phases at magic doping concentrations are the pristine electronic phases for high temperature superconductivity and further studies of the intrinsic phases will provide critical insights for the microscopic theory of HTS.

We would like to thank Zheng Wu (TcSUH), Logan Lebanowski (TcSUH), Jiwu Xiong (IOP) and Wenxin Ti (IOP) for their technical assistances. The work in Houston is supported by the State of Texas through The Texas Center for Superconductivity at the University of Houston. The work in Beijing is supported by Natural Science Foundation of China (NSFC) and Chinese Academy of Sciences (CAS). This work is also supported from CAS under the project: International Team on Superconductivity and Novel Electronic Materials (ITSNEM).

References

1.  S. H. Pan *et al.*, Nature **413**, 282 (2002).

2.  J. M. Tranquada *et al.*, Nature **375**, 561 (1995).

3.  Y. H. Kim, and P. H. Hor, Mod. Phys. Lett. B **15**, 497 (2001); P. H. Hor, and Y. H. Kim, J. Phys.: Condens. Matter **14**, 10377 (2002).

4.  Y. H. Kim *et al.*, J. Phys.: Condens. Matter **15**, 8485 (2003).

5.  M. Matsuda *et al.*, Phys. Rev. B **65**, 134515 (2002).

6.  B. Lorenz, Z.G. Li, T. Honma and P.-H. Hor, Phys. Rev. B **65**, 144522 (2002).

7. M. Vershinin *et al.*, Science **303**, 1995 (2004).

8. S. Komiya, H.D. Chen, S.C. Zhang and Y. Ando, Phys. Rev. Lett. **94**, 207004 (2005).

9. S. Sachdev and S. C. Zhang, Science **295**, 452 (2002).

10. F. Zhou *et al.*, Supercond. Sci. Technol. **16**, L7 (2003); F. Zhou *et al.*, Physica C **408-410**, 430 (2004).

11. J. A. Osborn, Phys. Rev **67**, 351 (1945).

12. We observed intrinsic $T_{C2}$ = 30 K at x = 0.11 (~1/9) consistent with a p(3x3) symmetry instead of previously suggested c(2x2) symmetry at x = 0.125 in reference 3. Currently the origin for this difference is unclear. However this will not affect our conclusions in this paper that robust intrinsic $T_C$'s occur at magic hole densities corresponding to 2D "square" electronic lattices.

13. All $T_C$'s reported from this point on will be the onset $T_C$ defined as the temperature where $d\chi/dT$ deviates from a linear baseline. $\Delta T_C$ is defined as the full width at the half maximum (FWHM) of $d\chi/dT$ curve, or $2(T_{FWHM} - T_{peak})$ when the lower part of $d\chi/dT$ peak is distorted due to the vortex phase transition mentioned in text.

14. T. Sasagawa *et al.*, Phys. Rev. B **61**, 1610 (2000).

15. R. Gilardi *et al.*, Euro. Phys. J. B **47**, 231(2005).

16. T. Matsushita *et al.*, Physica C **170**, 375 (1990). L. Krusin-Elbaum *et al.*, J. Appl. Phys. **67**, 4670, (1990).

17. Y. M. Huh *et al.*, Phys. Rev. B **63**, 064512 (2001).

18. See D. Mihailovic, Phys. Rev. Lett. **94**, 207001 (2005) and references therein.

19. B. Lake *et al.*, Nature Materials 4**, 658** (2005).

20. Y. H. Kim *et al.*, Phys. Rev. B **71**, 092508 (2005).

21. K. Krishana *et al.*, Science **277**, 83 (1997).

22. J. E. Sonier *et al.*, Phys. Rev. Lett. **83**, 4156 (1999).

**Figure 1:** Differential curves of FC susceptibility ($d\chi/dT$) vs. T in 5 Oe field along c-axis for SD-La214 single crystals of x = 0.063 (~1/16), 0.07, 0.09, 0.11 (~1/9) and 0.125 (=2/16).

**Figure 2:** FC $d\chi/dT$ curves under various fields for (a) 0.11 and (b) 0.125 SD-La214 crystals. As guided by solid black lines, the 30K SC phase of 0.11 crystal is 'robust' under fields up to 5 Tesla, on the other hand, the apparent initial $T_C$ (37 K) of 0.125 crystal decreases rapidly upon increasing H then stays at 30K for H ≥ 1 Tesla.

**Figure 3:** Field dependences of $T_C$ and transition width (color coded) for SD-La214 crystals with x of (a) 0.063 (~1/16), (b) 0.07, (c) 0.09, (d) 0.11 (~1/9) and (e) 0.125 (=2/16). Insets are $\Delta T_C$ vs. $T_C$ for 0.07 and 0.125 crystals, respectively.

**Figure 4:** Field dependences of $^{5K}\chi_{FC}$ for nine SD-La214 single crystals. Doping levels and masses are indicated. Also included are three sets of data points taken from literatures.

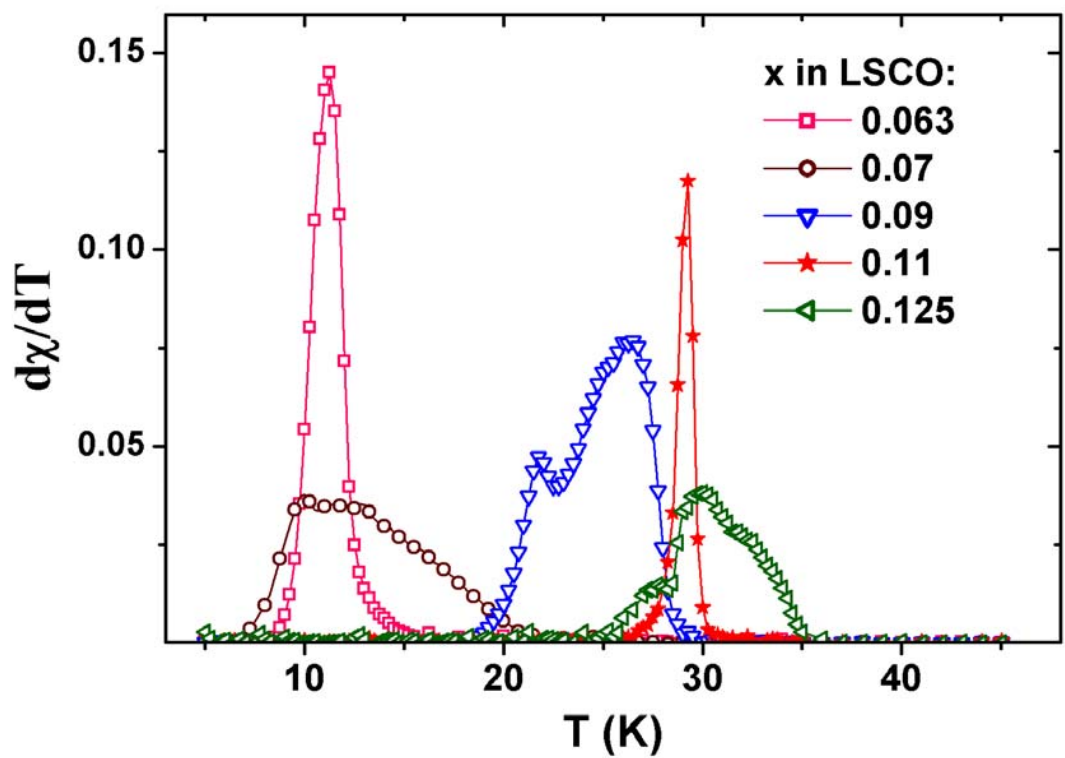

**Figure 1**

**Figure 2a**

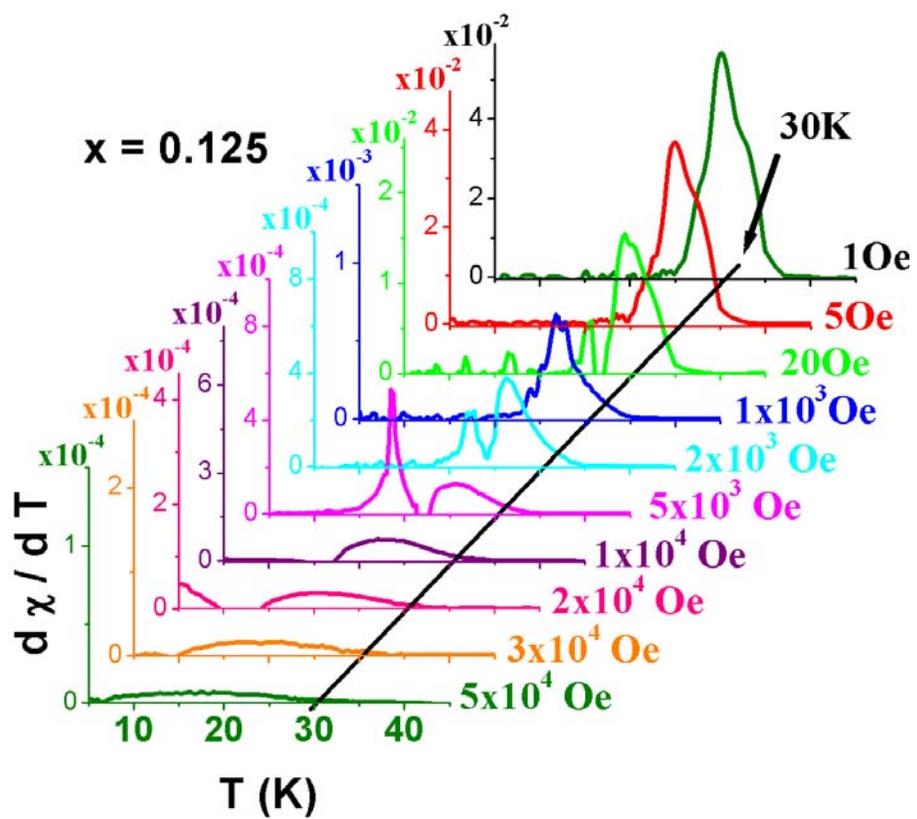

**Figure2b**

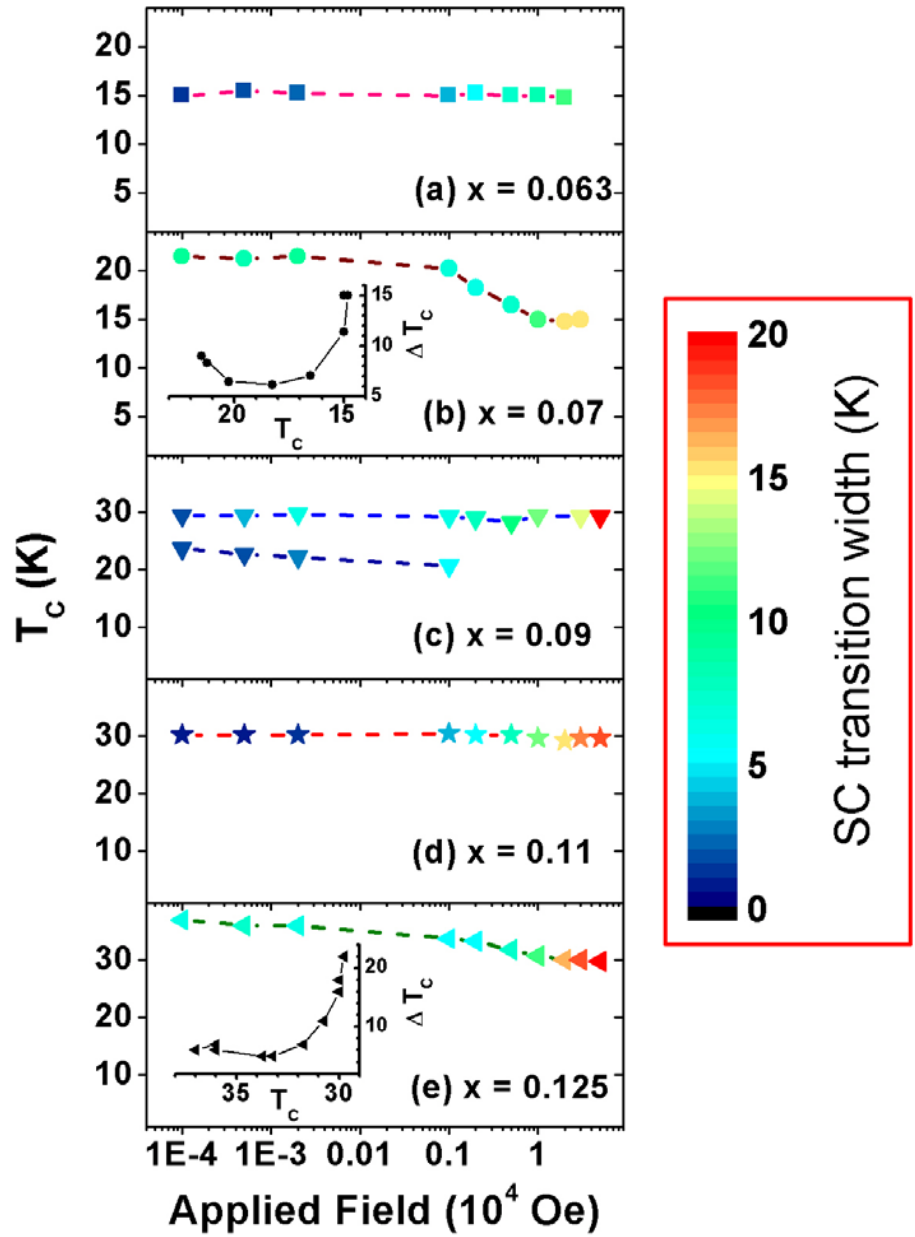

**Figure 3**

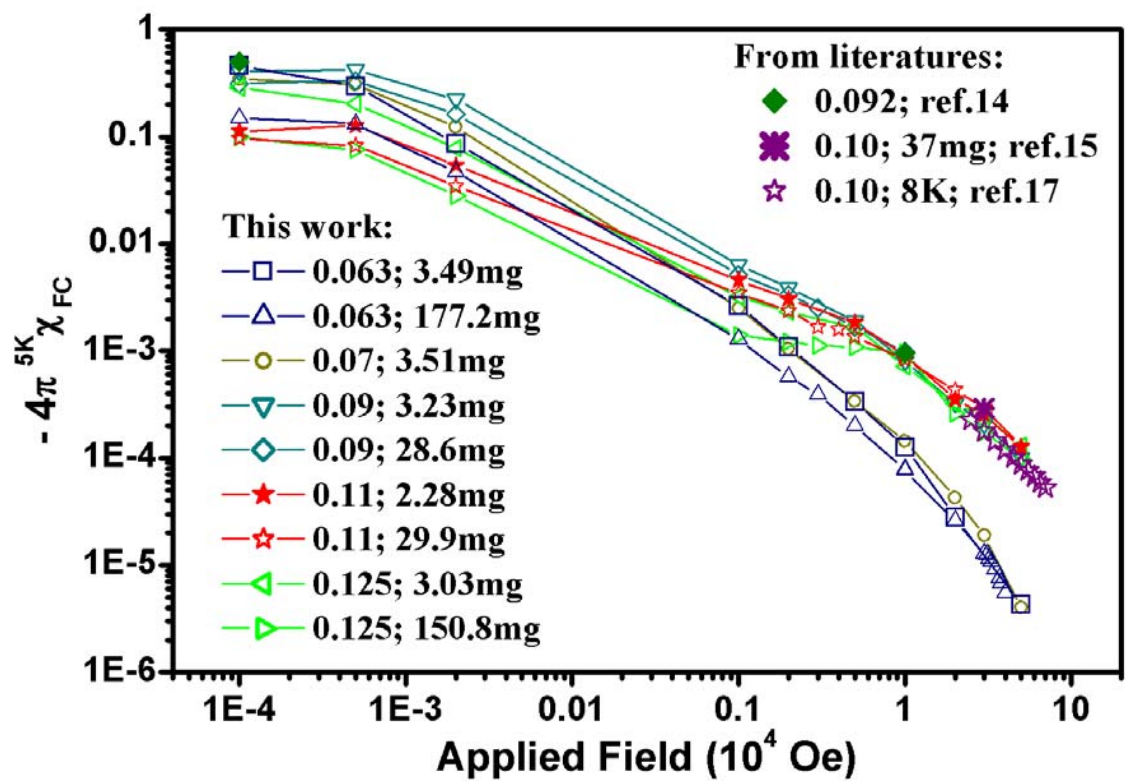

**Figure 4**